%Paper: hep-th/9511079
%From: Paul Townsend <P.K.Townsend@damtp.cam.ac.uk>
%Date: Sun, 12 Nov 95 21:51:36 GMT

%%%%%%%This requires the PHYZZX.TEX macropackage

%%%%%%%If you do not have the msbm fonts, delete the following 4 lines
\font\mybb=msbm10 at 12pt
\def\bb#1{\hbox{\mybb#1}}
\def\Z {\bb{Z}}
\def\R {\bb{R}}

%%%%%%%%%%%%
%%%and replace with the following 2 lines (without %)
%\def\Z {Z}
%\def\R {R}
%%%%%%%%%%

\tolerance=10000
\input phyzzx

 \def\unit{\hbox to 3.3pt{\hskip1.3pt \vrule height 7pt width .4pt \hskip.7pt
\vrule height 7.85pt width .4pt \kern-2.4pt
\hrulefill \kern-3pt
\raise 4pt\hbox{\char'40}}}

%%%%%%%%%%%%%%%%%%%%%%%%%%%%%%%%%%%%%%%%%%%%%%%%%%%%%%%%%%%%%%%%%%%%%%%%%%%%%
\REF\pol{ J. Polchinski, `Dirichlet-Branes and Ramond-Ramond charges',
hep-th/9510017.}
\REF\ggp{G. Gibbons, M.B. Green and M.J. Perry, {\it in preparation}.}
\REF\duff{M.J. Duff and P. van Nieuwenhuizen, Phys. Lett. {\bf 94B} (1980)
179.}
\REF\ant{A. Aurilia, H. Nicolai and P.K. Townsend, Nucl. Phys. {\bf B176}
(1980) 509.}
\REF\rom{L. Romans, Phys. Lett. {\bf 169B} (1986) 374.}
\REF\pols{ J. Polchinski and A, Strominger, `New Vacua for Type II String
Theory',
hep-th/9510227.}
\REF\Teit{M. Henneaux and C. Teitelboim, Phys. Lett. {\bf 143B} (1984) 415.}
\REF\PKT{P.K. Townsend, Phys. Lett. {\bf 350B} (1995) 184.}
\REF\EW{E. Witten, Nucl. Phys. {\bf B443} (1995) 85.}
\REF\JHS{J.H. Schwarz, {\it The Power of M-Theory}, hep-th/9510086.}
\REF\HW{P. Ho{\v r}ava and E. Witten, {\it Heterotic and Type I String
Dynamics from
Eleven Dimensions}, hep-th/9510209}.
\REF\PKTb{P.K. Townsend, {\it P-Brane Democracy}, hep-th/9507048, to appear in
the
proceedings of the March 1995 PASCOS/Johns Hopkins conference.}

%%%%%%%%%%%%%%%%%%%%%%%%%%%%%%%%%%%%%%%%%%%%%%%%%%%%%%%%%%%%%%%%%%%%

\Pubnum{ \vbox{ \hbox{R/95/55}  \hbox{hep-th/?} \hbox{UG-12/95} } }
\pubtype{}
\date{November, 1995}

\titlepage

\title {\bf The IIA super-eightbrane}

\author{E. Bergshoeff}
\address{Institute for Theoretical Physics,
\break
University of Groningen, Nijenborgh 4,
\break
9747 AG Groningen,
\break
The Netherlands}
\andauthor{M.B. Green, G. Papadopoulos and P.K. Townsend}
\address{DAMTP, University of Cambridge,
\break
Silver St., Cambridge CB3 9EW, U.K.}

\abstract{
We present a version of ten-dimensional IIA supergravity containing a 9-form
potential for which the field equations are equivalent to those of the
standard,
massless, IIA theory for vanishing 10-form field strength, $F_{10}$, and to
those of
the `massive' IIA theory for non-vanishing $F_{10}$. We exhibit a multi 8-brane
solution
of this theory which preserves half the supersymmetry. We propose this solution
as the
effective field theory realization of the Dirichlet 8-brane of type IIA
superstring
theory.}

\endpage
%\pagenumber=1

%%%%%%%%%%%%%%%%%%%%%%%%%%%%%%%%%CHAPTER 1%%%%%%%%%%%%%%%%%%%%%%%%%%%%%%%%%%%

\chapter{Introduction}

Recent advances in our understanding of non-perturbative superstring theory
have led to
the establishment of many connections between hitherto unrelated superstring
theories.
Many of these connections involve p-brane solutions of the respective
supergravity
theories that couple to the (p+1)-form potentials in the Ramond-Ramond (RR)
sector. These
RR p-branes are all singular as solutions of ten-dimensional (D=10)
supergravity, so their
status in superstring theory was unclear until recently. It now appears that
the RR
p-branes of type II supergravity theories have their place in type II
superstring theory
as `Dirichlet-branes', or `D-branes' [\pol]. These include the p-branes for
$p=0,2,4,6$ in
the type IIA case and the p-branes for $p=1,3,5$ in the type IIB case. However,
they also
include a type IIB (-1)-brane (instanton) and a 7-brane [\ggp], and a
type IIA 8-brane.

Since p-branes couple naturally to (p+1)-form potentials, the existence of an
8-brane in
type IIA D=10 superstring theory suggests the existence of a corresponding
9-form
potential, $A_9$, with 10-form field strength, $F_{10}$, in the effective type
IIA
supergravity theory. Assuming a standard kinetic term of the form $F_{10}^2$,
the
inclusion of this field does not lead to any additional degrees of freedom (per
spacetime
point) and so is not immediately ruled out by supersymmetry considerations, but
it allows
the introduction of a cosmological constant [\pol], as
explained many years ago in the context of a four-form field strength in
four-dimensional
field theories [\duff, \ant]. As it happens, a version of type IIA supergravity
theory
with a cosmological constant was constructed some time ago by Romans [\rom],
who called it
the `massive' IIA supergravity theory. This theory has the peculiarity that
D=10
Minkowski spacetime is {\it not} a solution of the field equations (and neither
is the
product of D=4 Minkowski spacetime with a Calabi-Yau space). Various
Kaluza-Klein (KK)
type solutions were found by Romans but none of them were supersymmetric, i.e.
these
solutions break all the supersymmetries.

Here we shall present the 10-form reformulation of Romans' theory. The new IIA
supergravity theory has the advantage that its solutions include those of both
the
massless and the massive IIA theory. We propose this new IIA supergravity
theory as the
effective field theory of the type IIA superstring, allowing for the
9-form potential. It has been suggested [\pol,\pols] that the expectation value
of
the dual of the 10-form field strength of this superstring theory should be
interpreted
as the cosmological constant of the massive IIA supergravity theory. One result
of this
paper is the determination of the precise relation between these quantities;
they are
conjugate variables in a sense discussed previously in the D=4 context [\Teit].
Our main result is the construction of multi 8-brane solutions of
the new IIA supergravity theory which preserve half the supersymmetry. These
solutions are
singular at the `centres' of the metric, but this is a general feature of RR
p-branes. We
propose these solutions as the effective field theory realization of Dirichlet
8-branes
of type IIA superstring theory.

We begin with a review of the massive IIA supergravity,
introducing some simplifications. We then construct the new formulation of the
bosonic
sector of this theory, incorporating the 9-form gauge field $A_9$, in which the
cosmological constant $m$ emerges as an integration constant. We then construct
a
supersymmetric 8-brane solution of the massive IIA supergravity theory and show
that it
has a generalization to multi 8-brane solutions of the new IIA theory. The
latter
solutions include some which are asymptotically flat. We shall comment further
on the
relation to type IIA superstring theory in the conclusions.

%%%%%%%%%%%%%%%%%%%%%%%%%%%%%%%%%CHAPTER 2%%%%%%%%%%%%%%%%%%%%%%%%%%%%%%%%%%%

\chapter{The massive IIA supergravity}

The bosonic field content of the massive IIA D=10 supergravity theory comprises
(in our notation) the (Einstein) metric, $g^{(E)}$, the dilaton, $\sigma$, a
massive
2-form tensor field
$B'$ and a three-form potential $C'$. One introduces the field-strengths
$$
\eqalign{
G &=4dC' + 6m(B')^2\cr
H &=3dB' }
\eqn\aone
$$
where $m$ is a mass parameter. The Lagrangian for these fields is [\rom]
$$
\eqalign{
{\cal L} &= \sqrt{-g^{(E)}}\; \Big[ R_{(E)} -{1\over2}|\partial\sigma|^2 -
{1\over3}e^{-\sigma}|H|^2 -{1\over12}e^{{1\over2}\sigma}|G|^2 -
m^2e^{{3\over2}\sigma}|B'|^2
-{1\over2}m^2 e^{{5\over2}\sigma}\Big]\cr
& + {1\over 9}\varepsilon \big[ dC'dC'B' + mdC'(B')^3 + {9\over20}m^2
(B')^5\big]\ .}
\eqn\atwo
$$
The notation for forms being used here is that a $q$-form $Q$ has components
$Q_{M_1\dots M_q}$ given by
$$
Q= Q_{M_1\dots M_q}dx^{M_1}\wedge \dots \wedge dx^{M_q}\ .
\eqn\athree
$$
Thus, the $(1/9)\varepsilon dC'dC'B'$ term in \atwo\ is shorthand for
$$
{1\over 9}\varepsilon^{M_1\dots
M_{10}}\partial_{M_1}C'_{M_2M_3M_4}\partial_{M_5}C'_{M_6M_7M_8}B_{M_9M_{10}}\ .
\eqn\aathree
$$

As explained in [\rom] the massless limit is not found by simply setting $m=0$
in \atwo\
because the supersymmetry transformations involve terms containing $m^{-1}$.
Instead,
one first makes the field redefinitions
$$
\eqalign{
B' &= B + {2\over m}dA\cr
C' &= \tilde C - {6\over m} AdA\ . }
\eqn\afour
$$
This redefinition introduces the gauge invariance
$$
\eqalign{
\delta A &= -m\Lambda \cr
\delta B &= 2d\Lambda\cr
\delta\tilde C &= 12 Ad\Lambda}
\eqn\afive
$$
for which the gauge-invariant field strengths are
$$
\eqalign{
F&= 2dA + mB\cr
H&= 3dB\cr
G &= 4d\tilde C + 24 BdA + 6m B^2\ .}
\eqn\asix
$$
The bosonic Lagrangian of the massive IIA theory is now
$$
\eqalign{
{\cal L} &= \sqrt{-g^{(E)}}\; \Big[ R_{(E)} -{1\over2}|\partial\sigma|^2 -
{1\over3}e^{-\sigma}|H|^2 -{1\over12}e^{{1\over2}\sigma}|G|^2 -
e^{{3\over2}\sigma}|F|^2
-{1\over2}m^2 e^{{5\over2}\sigma}\Big]\cr
& + {1\over 9}\varepsilon \big[ d\tilde C d\tilde C B + 6d\tilde C B^2 dA +
12(dA)^2B^3 +
md\tilde C B^3 + {9\over2}mB^4 dA + {9\over 20}m^2 (B)^5\big]\ ,}
\eqn\aseven
$$
and the bosonic Lagrangian of the massless IIA theory can now be found by
taking the
$m\rightarrow 0$ limit.

The Lagrangian \aseven\ can be simplified by the further redefinition
$$
\tilde C = C-6AB\ .
\eqn\aeight
$$
The $\Lambda$-gauge transformation of the new 3-form $C$ is
$$
\delta C = -6m\Lambda B
\eqn\anine
$$
and the gauge-invariant field strengths, $F$, $H$, and $G$ are now given by
$$
\eqalign{
F&= 2dA + mB\cr
H&= 3dB\cr
G &=4dC + 24AdB + 6mB^2 \ .}
\eqn\aten
$$
At the same time, to make contact with string theory, it is convenient to
introduce the
string metric
$$
g_{MN} = e^{-{1\over2}\sigma}g_{MN}^{(E)}\ .
\eqn\aeleven
$$
The bosonic Lagrangian now takes the simple form
$$
\eqalign{
{\cal L} &= \sqrt{-g}\big\{e^{-2\sigma}\; \big[ R -{1\over2}|\partial\sigma|^2
-
{1\over3}|H|^2\big]  - |F|^2 - {1\over12}|G|^2
- {1\over2}m^2\big\}\cr
& + {1\over 9}\varepsilon\big[ dC dC B + mdCB^3  + {9\over 20}m^2 (B)^5\big]\
.}
\eqn\atwelve
$$
Observe that the final topological term is simply a type of Chern-Simons (CS)
term
associated with the 11-form $G^2H$. Thus, the bosonic action of the massive
type IIA
supergravity theory can be written as
$$
\eqalign{
I = \int_{{\cal M}_{10}}\! d^{10}x\; &\sqrt{-g}\Big\{ e^{-2\sigma}\; \Big[ R
-{1\over2}|\partial\sigma|^2 - {1\over3}|H|^2\Big]  - |F|^2 -
{1\over12}|G|^2 - {1\over2}m^2\Big\} \cr
& + {1\over 9}\int_{{\cal M}_{11}}\!  G^2 H \ ,}
\eqn\athirteen
$$
where ${\cal M}_{11}$ is an 11-manifold with boundary ${\cal M}_{10}$. Apart
from
the cosmological constant, the $m$-dependent terms in the action can be simply
understood as arising from the replacement of the usual $m$-independent field
strengths
of the massless type IIA theory by their $m$-dependent generalizations \aten.
Furthermore, the $m$-dependence of these field strengths is
completely fixed by the `shift' gauge transformation $\delta A=-m\Lambda$ of
$A$, as are
the $\Lambda$-gauge transformations. The relation of the constant $m$ appearing
in
this transformation with the cosmological constant cannot be understood purely
within the
context of the bosonic Lagrangian but is, of course, fixed by supersymmetry.

Observe that the cosmological constant term in \athirteen\ is now (in the
string
metric) independent of the dilaton. This is typical of the RR sector and is
consistent
with the idea that $m$ can be interpreted as the expectation value of the dual
of a RR
10-form field strength. This interpretation would have the additional virtue of
restoring
the invariance under the discrete symmetry in which all RR fields change sign,
a symmetry
that is broken by the terms linear in $m$ in \atwelve. We shall now show how to
reformulate the massive IIA theory along these lines. As we shall see the
cosmological
constant is simply related to, but not equal to, the expectation value of the
ten-form
field strength.

%%%%%%%%%%%%%%%%%%%%%%%%%%%%%%%%%CHAPTER 3%%%%%%%%%%%%%%%%%%%%%%%%%%%%%%%%%%%

\chapter{IIA supergravity with 9-form potential}

We shall start with the bosonic Lagrangian of \atwelve. Expanding in powers of
$m$, the
associated action $I(m)$ is
$$
\eqalign{
I(m) &= I(0) + \int \! d^{10} x \;  \Big\{ 2m\sqrt{-g}\big[(dC+6AdB)\cdot B^2
-2dA\cdot
B\big] + {m\over 9}\varepsilon dCB^3\big]\cr
\qquad & -{1\over2}m^2\sqrt{-g}\Big[1+2|B|^2+6|B^2|^2\Big] +{m^2\over
20}\varepsilon
B^5\Big]\Big\} \ ,}
\eqn\none
$$
where $I(0)$ is the bosonic action of the massless IIA supergravity theory.
We now promote the constant $m$ to a field $M(x)$, at the same time introducing
a 9-form
potential $A_9$ as a Lagrange multipler for the constraint $dM=0$. Omitting a
surface
term, the Lagrange multiplier term can be rewritten as
$$
10\; \varepsilon dA_9 M\ .
\eqn\ntwo
$$
The $A_9$ field equation implies that $M=m$, for some constant
$m$, so the the remaining equations are equivalent to those of the massive IIA
theory
except that the constant $m$ is now arbitrary and that we now have an
additional field equation from varying $M$. This additional equation is
$$
{\delta I(M) \over \delta M(x)} =  -\varepsilon F_{10}
\eqn\nthree
$$
where $I(M)$ is the action \none\ but with $M$ replacing $m$, and $F_{10}= 10\;
dA_9$ is
the 10-form field strength of $A_9$. Thus the $M$ equation simply determines
the new field
strength $F_{10}$.  Observe that the expectation value of $(\varepsilon
F_{10})$ is {\it
not} equal to the expectation value of $\sqrt{-g}M$, as a matter of principle
(although it
may equal it in special backgrounds) , but is rather the value of the variable
canonically
conjugate to it.

Note that the gauge and supersymmetry transformations of the action $I(M)$ no
longer
vanish. However, the variations of $I(M)$ are proportional to $dM$ and can
therefore be
cancelled by a variation of the new 9-form gauge potential $A_9$. This
determines the
gauge and supersymmetry transformations of $A_9$. The supersymmetry variation
will not be
needed for our purposes so we omit it. The $\Lambda$-gauge transformation of
$A_9$
found in this way is
$$
\delta (\varepsilon A_9) = {2\over5}\sqrt{-g} \Big[\Lambda\cdot F + (\Lambda
B)\cdot
G\Big] - {1\over30}\varepsilon \big( 2\Lambda dC B^2 + M\Lambda B^4\big)\ .
\eqn\aextra
$$
We now have a new gauge-invariant bosonic action
$$
I(M) + \int d^{10}x\; M\varepsilon F_{10}\ .
\eqn\nfour
$$
The field $M$ can now be treated as an auxiliary field that can be eliminated
via its
field equation
$$
\sqrt{-g}M = K^{-1}(B)\Big\{ \varepsilon (F_{10} + {1\over 9}dC B^3)
+2\sqrt{-g}[(dC + 6AdB)\cdot B^2 - 2dA\cdot B]\Big\}\ ,
\eqn\nfive
$$
where
$$
K(B) =  1+2|B^2|+6|B^2|^2-{1\over10\sqrt{-g}}\varepsilon B^5\ .
\eqn\nsix
$$
Using this relation in \nfour\ we arrive at the Lagrangian
$$
{\cal L}_{new} = {\cal L}_0 + \big[\sqrt{-g}K(B)\big]^{-1} \Big\{ \varepsilon
(F_{10} +
{1\over 9}dC B^3) +2\sqrt{-g}[(dC + 6AdB)\cdot B^2 - 2dA\cdot B]\Big\}^2
\eqn\nseven
$$
where ${\cal L}_0$ is the bosonic Lagrangian of the massless IIA theory.
Note the non-polynomial structure of the new Lagrangian in the gauge field $B$.
This
greatly obscures the $\Lambda$-gauge invariance, which is ensured by the very
complicated $\Lambda$-gauge transformation of $A_9$.

%%%%%%%%%%%%%%%%%%%%%%%%%%%%%%%%%CHAPTER 4%%%%%%%%%%%%%%%%%%%%%%%%%%%%%%%%%%%

\chapter{The Eightbrane}

The appearance of the 9-form potential in the above reformulation of the
massive IIA
supergravity theory suggests the existence of an associated 8-brane solution.
We will
find solutions of the equations of motion of \nseven\ of the form
$$
\eqalign{
ds_{(E)}^2 &= f^2(y)\; dx^\mu dx^\nu\eta_{\mu\nu} + dy^2\cr
\sigma &= \sigma(y) \cr
A_9 &= A_9(y)}
\eqn\bone
$$
with all other fields vanishing, and where $\eta$ is the Minkowski 9-metric.
Such a
solution will have 9-dimensional Poincar\'e invariance and hence an
interpretation as an
8-brane. We shall further require of such a solution that it preserve some
supersymmetry,
so we shall begin by considering the variation of the gravitino one-form $\psi$
and the
dilatino $\lambda$ in the presence of configurations of the above form. The
full
variations of the massive IIA theory can be found in [\rom]. They depend on the
constant
$m$. In the new theory, this constant is replaced by the function $M$ given in
\nfive.
For the backgrounds considered here, $\sqrt{-g}M= \varepsilon F_{10}$ and the
supersymmetry variations of the fermions reduce to
$$
\eqalign{
\delta_\epsilon \psi &= D\epsilon -{1\over 32}M
e^{{5\over4}\sigma}\Gamma\epsilon\cr
\delta_\epsilon \lambda &= -{1\over 2\sqrt{2}}\big(\Gamma^M\partial_M\sigma +
{5\over4}Me^{{5\over4}\sigma}\Big)\epsilon\ . }
\eqn\btwo
$$

For configurations of the assumed form and further assuming that $\epsilon$
depends only
on $y$, the equations $\delta\psi=0$ and $\delta\lambda=0$ become
$$
\eqalign{
0&=\epsilon' -{1\over 32}M e^{{5\over4}\sigma} \Gamma_y\epsilon
\cr
0& = \big( f'\Gamma_y
-{1\over 16}M fe^{{5\over4}\sigma} \big) \epsilon \cr
0&= \big(\sigma' + {5\over4}Me^{{5\over4}\sigma}\Gamma_y\big)\epsilon \ ,}
\eqn\bthree
$$
where the prime indicates differentiation with respect to $y$.
To find non-zero solutions for $\epsilon$ we are now forced to suppose that
$$
\Gamma_y\epsilon =\pm \epsilon\ .
\eqn\bfour
$$
We then find that
$$
f' = \pm {1\over16}Mfe^{{5\over4}\sigma}
\eqn\abfive
$$
and that
$$
\Big(e^{-{5\over4}\sigma}\Big)' =\pm {25\over 16}M\ .
\eqn\bfive
$$
Eliminating $M$ from these equations we deduce that
$$
f= Ae^{-{1\over20}\sigma}\ ,
\eqn\bsix
$$
for some constant $A$. Using now the $A_9$ field
equation $M'=0$, \bfive\ is seen to imply that
$$
\partial^2_y\Big(e^{-{5\over4}\sigma}\Big)=0\ .
\eqn\bsixa
$$
The general solution is given in terms of a harmonic function $V(y)$, the
precise nature
of which will be discussed shortly, i.e.
$$
e^{-{5\over4}\sigma}= V(y)\ .
\eqn\bseven
$$
Equation \bfive, together with \bsix, now gives
$$
M = \mp {16\over25} V'\ ,\qquad\qquad f=AV^{1\over25} \ .
\eqn\beight
$$
In principle, we have still to consider the other field equations, but we have
checked
that they are all solved by the above field configurations.
The Einstein metric of the multi 8-brane solution is
$$
ds^2_{(E)} = A^2V^{2\over25}(y) dx^\mu dx^\nu\eta_{\mu\nu} + dy^2\ .
\eqn\bnine
$$
The string metric is
$$
ds^2= A^2V^{12\over25}dx^\mu dx^\nu\eta_{\mu\nu} + V^{2\over5}dy^2\ .
\eqn\bten
$$
The Killing spinor
$\epsilon$ is given by
$$
\epsilon = V^{1\over 50}\epsilon_0,\qquad \Gamma_y\epsilon_0 =\pm \epsilon_0\ ,
\eqn\bseven
$$
where $\epsilon_0$ is a constant spinor.

It remains to determine the function $V$. Consider first the massive IIA theory
for which
the function $M$ equals the (non-zero) constant $m$ appearing in the
Lagrangian, which we
may choose to be positive. In this case
$$
V=\pm {25\over16}m(y-y_0)
\eqn\gone
$$
where the sign depends on the choice of chirality of $\epsilon$. However, $V$
must be
positive for real $\sigma$, so the spinor $\epsilon$ must change chirality at
$y=y_0$.
This is possible because the spinor $\epsilon$ vanishes at $y=y_0$. This is
acceptable
because the metric (either the Einstein or the string one) is also singular
at $y=y_0$. Thus, the massive IIA theory has a solution for
which
$$
V= {25\over16}m|y-y_0|\ .
\eqn\gtwo
$$
Note that $V$ is a continuous function of $y$ with a kink singularity at
$y=y_0$, at
which the curvature tensor has a delta function singularity.

In the new IIA theory we may suppose that $M$ is only {\it locally} constant.
The form
of the function $V(y)$ in this case depends on the type of point singularity
that we
allow. The above example suggests that we should require $V$ to be a continuous
function
of $y$. There are solutions with discontinuities in $V$ but they have $\delta'$
type
singularities of the curvature tensor, and we shall not consider them. In any
case, the
restriction to kink singularities produces physically sensible results, as we
shall see.
An example of a solution with a single kink singularity of $V$ is
$$
V=\cases{-ay+b \qquad y<0\cr cy+b\qquad y>0}
\eqn\gthree
$$
where $a$, $b$ and $c$ are non-negative constants. We adopt this as the basic
single
8-brane solution. It can be interpreted as a domain wall separating regions
with
different values of $M$. The solution has two asymptotic regions relative to
which an
8-brane charge, $Q_\pm$, may be defined as the value of $M$ as $y\rightarrow
\pm \infty$.
For the above solution,
$$
Q_+ =c\qquad Q_-= a
\eqn\gfour
$$
The multi 8-brane generalization of \gthree\ with the same charges is found by
allowing
kink singularities of V at n+1 ordered points $y=y_0<y_1<y_2<\dots <y_n$.
The function $V$ is
$$
V=\cases{-a(y-y_0) + \sum_{i=1}^n\mu_i (y_i-y_0) + b\qquad y<y_0 \cr
 (c-\sum_{i=1}^n\mu_i) |y-y_0| + \sum_{i=1}^n \mu_i |y-y_i| +b \qquad y>y_0}
\eqn\gfive
$$
where $\mu_i$ are positive constants and $a$, $b$, $c$ are non-negative
constants.

The asymptotically left-flat or right-flat solutions are those for which
$Q_-=0$ or
$Q_+=0$, respectively. The asymptotically flat solutions are those which are
both
asymptotically left-flat and right-flat. An example of an asymptotically flat
three
8-brane solution is given by $V= \mu^2\big||y-y_0|-|y-y_1|\big| + \gamma^2$,
where $\mu$ and $\gamma$ are arbitrary constants.

 %%%%%%%%%%%%%%%%%%%%%%%%%%%%%%%%%CHAPTER 4%%%%%%%%%%%%%%%%%%%%%%%%%%%%%%%%%%%

\chapter{Comments}

One obvious point to be considered is the nature of the nine-dimensional
worldvolume
field theory governing the dynamics of small perturbations about the static
single 8-brane solution. Since the solution preserves half the D=10 type II
supersymmetry,
this worldvolume field theory must have N=1 nine-dimensional supersymmetry. It
must also
include one Nambu-Goldstone scalar field corresponding to the breaking of
translational
invariance in the $y$ direction. Given that the worldvolume action does not
include fields
of spin greater than one, there is a unique candidate that fulfils these
requirements,
namely the nine-dimensional super-Maxwell multiplet.
This result is suggestive. Note that the 8-brane solution \bnine\ can be
double-dimensionally reduced to yield a membrane in a D=4 N=8 theory (by
periodic
identification of six spatial coordinates). Given that the worldvolume fields
of the
membrane are those of a nine-dimensional super-Maxwell multiplet, the
worldvolume
fields of the D=4 membrane will be those of an N=8 three-dimensional
super-Maxwell
multiplet, which is equivalent by dualization of the vector field to the
worldvolume
supermultiplet of the D=11 supermembrane. This suggests a possible connection
of the
massive IIA supergravity to D=11 supergravity via the reduction of the D=11
supermembrane
to D=4.  As a solution of D=11 supergravity, the latter is also determined in
terms of a
single harmonic function $V$. Normally $V$ is harmonic on the transverse
eight-dimensional
Euclidean space, but after compactification on $T^7$ to D=4 we need a harmonic
function on
$\R\times T^7$. The solution that is constant on $T^7$ is therefore similar to
that given
above for the type II 8-brane and we have verified that it is a supersymmetric
solution
of the massive N=8 D=4 supergravity theory constructed in [\ant], so its status
is rather
similar to that of the 8-brane solution of the massive type IIA supergravity.

The single 8-brane solution described in this paper should be related to the
Dirichlet
8-brane of [\pol]. This is a string background in which open string states
arise with
fixed (Dirichlet) boundary conditions that are imposed in one space-like
dimension at
one or both ends of the string. These conditions restrict at least one of the
end-points
of open strings to lie in the nine-dimensional worldvolume of an 8-brane. The
8-brane
couples to a 9-form gauge field with a ten-form field strength $F_{10}$. If the
new IIA
supergravity constructed here is indeed the effective field theory of the IIA
superstring
in the presence of this 10-form field strength then it should be possible to
recover the
Lagrangian \nseven\ by string theory considerations. Neglecting terms of order
$B^2$,
which in any case follow from gauge invariance, the only term in \nseven\ that
is linear
in $F_{10}$ is proportional to
$$
(\varepsilon F_{10}) dA\cdot B\ .
\eqn\kone
$$
This is the crucial term that has to be reproduced in string theory.
There is a vertex operator in the RR sector of the type IIA theory that
couples a ten-form field strength to the worldsheet. This vertex operator has
the form
$F_{10}{\bar S}S$, where $S$ is the spacetime spinor worldsheet field of the
spacetime
supersymmetric worldsheet action. There are non-trivial tree diagrams that mix
$F_{10}$
with fields from the RR and NSNS sectors, producing a term of the form \kone,
as
required. The requirements of gauge invariance suggest that a more systematic
consideration of string theory in the presence of D-branes would produce the
full
effective Lagrangian \nseven.

Since all the $p$-brane solutions of D=10 IIA supergravity for $p<8$ can be
viewed as
arising from some 11-dimensional `M-theory' [\PKT,\EW,\JHS,\HW], it would be
surprising
if the 8-brane did not also have an 11-dimensional interpretation. The obvious
possibility is that the D=10 8-brane is the double-dimensional reduction of a
D=11
supersymmetric 9-brane. Such an object would be expected (see [\PKTb]) to carry
a 9-form
`charge' appearing in the D=11 supertranslation algebra as a central charge.
This is
possible because the 2-form charge normally associated with the D=11
supermembrane is
algebraically equivalent to a 9-form. It is not easy to see how to implement
this idea,
however, since there is no `massive' D=11 supergravity theory. One possibility
is
suggested by the recent interpretation [\HW] of the heterotic string as an
$S^1/\Z_2$
compactified M-theory. Since the compactification breaks half the supersymmetry
and the
compactifying space is actually the closed interval, the two D=10 spacetime
boundaries
might be viewed as the worldvolumes of two D=11 9-branes.

\vskip 1cm
\centerline{\bf Acknowledgements}
\vskip 0.5cm
G.P. is supported by a University Research Fellowship from the Royal Society.
One of us
(E.B.) would like to thank DAMTP for its hospitality.  The work of E.B. has
been made
possible by a fellowship of the Royal Netherlands Academy of Arts and Sciences
(KNAW).

\refout

\end